\documentclass[conference, 10pt, compsocconf]{IEEEtran}
\ifCLASSOPTIONcompsoc
  \usepackage[nocompress]{cite}
\else
  \usepackage{cite}
\fi

\ifCLASSINFOpdf
\else
\fi

\usepackage[utf8]{inputenc}
\usepackage{url}
\usepackage{enumitem}
\usepackage{graphicx}
\usepackage[english]{babel}
\usepackage{csquotes}
\usepackage{amsmath}
\usepackage{amssymb}

\DeclareMathOperator*{\mean}{mean}
\DeclareMathOperator*{\median}{median}
\DeclareMathOperator*{\std}{std}

\begin{document}

\title{RadIoT: Radio Communications Intrusion Detection for IoT - A Protocol Independent Approach}

\author{\IEEEauthorblockN{Jonathan Roux, Éric Alata,
Guillaume Auriol, Mohamed Kaâniche, Vincent Nicomette, Romain Cayre}
\IEEEauthorblockA{LAAS-CNRS, Université de Toulouse, CNRS, INSA, Toulouse, France\\
Email : \{jonathan.roux, eric.alata, guillaume.auriol, mohamed.kaaniche, vincent.nicomette,
romain.cayre\}@laas.fr}
}

\maketitle

\begin{abstract}

    Internet-of-Things (IoT) devices are nowadays massively integrated in daily life: homes, factories, or public places. This technology offers attractive services to improve the quality of life as well as new economic markets through the exploitation of the collected data. However, these connected objects have also become attractive targets for attackers because their current security design is often weak or flawed, as illustrated by several vulnerabilities such as Mirai, Blueborne, etc.
    This paper presents a novel approach for detecting intrusions in smart spaces such as smarthomes, or smartfactories, that is based on the monitoring and profiling of radio communications at the physical layer using machine learning techniques. The approach is designed to be independent of the large and heterogeneous set of wireless communication protocols typically implemented by connected objects such as WiFi, Bluetooth, Zigbee, Bluetooth-Low-Energy (BLE) or proprietary communication protocols. The main concepts of the proposed approach are presented together with an experimental case study illustrating its feasibility based on data collected during the deployment of the intrusion detection approach in a smart home under real-life conditions.

\end{abstract}


\IEEEpeerreviewmaketitle

\section{Introduction}
\label{chap:intro}

The Internet of Things (IoT) has received increasing interest in the last few
years. Indeed, smart connected objects are nowadays widely used in our daily life activities, at home as well as in public and
professional spaces. Many of such objects (speakers, TVs, cameras,
doors, shutters, lightbulbs, sensors, etc.), that used to lack connectivity
in the past, can nowadays interact with other objects nearby or remotely through the Internet, using heterogeneous communication
protocols, such as Zigbee, Bluetooth or even non-standardized ones. While such
evolution enables the development of attractive services for the users,
serious concerns can be raised with respect to the new opportunities offered by
IoT objects to attackers to threaten the security and privacy of the users \cite{kolias_learning_2016}.

Pervasive connectivity is more and more used to manage more efficiently the home energy consumption
\cite{Wayes_2018}, the physical security of house
appliances \cite{Kokkonis2017}, or just for entertainment. Hence, a
smarthome has become nowadays a quite complex environment, composed of heterogeneous connected
objects, including mobile devices, that may be compliant or not
to standard communication protocols. Such smart objects generally belong to different
users living in the smart home, or to some visitors who bring
their own personal devices. Also, several communication protocols with
different characteristics (channel frequency, bandwidth, etc.) are generally used in the same connected environment. The objects may interact through gateways but may also communicate directly through ad hoc protocols. In the same time, thanks to the ability
of the connected objects to be programmed to automatically execute some predefined tasks, using e.g., new emerging devices like home assistants (Google Home, Amazon Echo, …), the
communication patterns resulting from the activities carried out in smarthomes 
become more and more deterministic, at least at some periods of the day \cite{IFTTT}. For instance, one
may wish to program automatically the opening of his connected shutters each
morning at a specific moment and may wish to program his connected coffee
machine at the same time. This trend is likely to be amplified in the
coming years.

The large majority of smart object users are not security experts and are not aware of the vulnerabilities and the security threats induced by their network-accessible
objects. Therefore, smarthomes become relevant targets for attackers, for different objectives. First, as some objects
are often used to collect personal information about the environment, the privacy
of users can be threatened if one of them is compromised. Furthermore, many objects
are currently used to protect the physical access to the house, e.g., camera,
locks, blinds, and an attacker may try to corrupt them to enter the house without
causing any damage \cite{dhanjani_abusing_2015}. The recent massive attack Mirai
\cite{MIRAI} shows that connected devices with weak security
protections may also be compromised in order to be enrolled in a botnet. 

The traditional IT security protections, such as firewalls or network intrusion detection
systems (IDS) usually rely on a central device, such as a gateway or
a proxy, to analyze the network communications, in order to
block unauthorized communications or 
to raise alarms. However, such solutions have generally a limited scope. In particular, direct  communications between objects using ad hoc or proprietary protocols are generally not covered. Also, some existing solutions consist in the development of specific monitors covering the communications associated to only a few and already known specific wireless protocols (in particular WiFi). Such solutions are not efficient to cover the large set of protocols used by the variety of connected objects deployed in a smart home. Also, they are not easy to adapt to take into account new objects or communication protocols integrated in the smarthome. 

The objective of this paper is to propose a novel approach to detect attacks
targeting a smarthome that addresses the shortcomings of existing solutions. Our intrusion detection approach, called  RadIoT, consists in the monitoring of the radio communication activities generated by the connected objects and the profiling of such activities based on machine learning techniques. The advantages of operating at the physical layer level are to be: 1) independent
of the different communication protocols used so far, and to take into account new
communication protocols that may emerge, 2) 
transparent from the objects point of view, i.e., the approach does not require the
modification of the connected objects, and 3) easily adaptable to any
smarthome network configuration modification (for example, addition or removal of new objects).

The implementation of the proposed approach is based on the deployment in a smarthome of a set of probes to capture the radio communication activities on predefined frequency bandwidths, using \textit{Software Defined Radio (SDR)} technology, and the elaboration of a set of models using machine learning techniques that reflect the radio activities recorded in a smart home without attacks during different periods of the day, the week, or during vacations. Each deviation from these models is considered as a potential malicious attack and raises an alert. The models can take into account malicious activities that
possibly involve the collaboration of several smart objects.

This paper describes the main concepts behind the proposed approach and presents some preliminary experiments aimed at assessing its feasibility in a real environment. It is organized as follows. Section \ref{chap:related} first presents a
state of the art of IoT attacks and network solutions to detect them. Section \ref{chap:threat} 
explains the threat model considered, as well as examples of malicious behaviors 
that our IDS is able to detect. Section \ref{chap:overview} presents the main concepts 
behind our proposed intrusion detection system. Section \ref{chap:design} and Section \ref{chap:model} present
respectively the architecture of the solution and how the model is generated. Next,
Section \ref{chap:expe} describes the experimentation carried out and the detection
results of our approach. Finally, Section \ref{chap:limits} discusses the
limitations and the future work to improve our IDS. 
\section{Related work}
\label{chap:related}

This section focuses first on smarthomes related attacks
targeting IoT devices, then on some security mechanisms proposed in related work 
to cope with such attacks. 

In the recent years, several attacks based on traditional Internet protocols exploiting weaknesses in IoT devices used in smarthomes have been reported \cite{dhanjani_abusing_2015}. A popular example is the Mirai worm. 
The aim of Mirai was to take the control of many connected objects in order to
create a botnet, that is then used to attack the DNS servers of
some famous companies like Amazon and Facebook \cite{MIRAI}. One of the vulnerabilities exploited by Mirai was the presence of weak credentials configured by default during the initialization of the connected objects.

Other attacks exploit weaknesses in the short distance IoT
communication protocols, like Zigbee, Bluetooth or even in proprietary ones.
In \cite{ronen_extended_2016}, E. Ronen et al. show that knowing the master key of
the Philips Hue system, unfortunately shared between all gateways and lightbulbs, it is possible
to reset the connection between a gateway and a lightbulb and then control the lightbulb.
The attacker only needs to send one very-high powered message from outside of the building
to initiate a new connection with the lightbulbs. This
vulnerability was investigated to design a potential worm for lightbulbs \cite{Ronen_IoTGoesNuclear_2018}.

M. Ryan \cite{ryan_bluetooth_2013} presents a vulnerability in BLE devices allowing attackers to brute-force encryption keys. This vulnerability 
was used in a few technical demonstrations to show that an attacker can open doors
just by using a replay attack \cite{Jasek_GattackingBluetoothsmart_2016}.   
Another example is the BlueBorne attack \cite{Ben_BlueBorneTechnicalWhite_2017} that allows an attacker to exploit some weak implementations of Bluetooth
on most architectures (IoT, computers, smartphones) to control or extract
information remotely.
Some classifications of IoT vulnerability attacks have been also proposed. For example, the Open Web Application Security Project
(OWASP) \cite{owasp_owasp_2017} describes common vulnerabilities
of IoT devices like memory leak, authentication weakness, etc. In 
 \cite{ronen_extended_2016}, attacks are categorized according to their impact on the functionality of the device as
perceived by end-users.

Many existing security protection mechanisms for IoT focus on the traditional Internet protocols, e.g., WiFi and Ethernet. For example, IoT SENTINEL \cite{miettinen_iot_2016} probes WiFi and Ethernet 
traffic flows through the access point of the smarthome. The connected devices are
identified by their MAC address. The tool identifies the device type as well as its potential vulnerabilities reported
in the CVE database. Vulnerable IoT devices are isolated from the Internet and from the other ones using
filtering rules. Accordingly,
the device functionality maybe heavily impacted by the filtering mechanisms, and as the IoT objects cannot be easily patched, they could become useless. Also, it is noteworthy that the attacks carried out through other wireless protocols, like ZigBee or Bluetooth, are not covered by IoT SENTINEL. 

Some research works \cite{sung_intelligent_2016} investigated the possibility to use RSSI (Radio Signal Strength Indicator) to detect physical intrusions within a Bluetooth connected area. This indicator
is mostly used for indoor localization \cite{dang_improved_2016}. To our viewpoint, such 
detection approaches are promising, and, in this paper, we propose to extend them to detect logical intrusions
on any protocols, based on the monitoring of radio communications. 

To conclude, current solutions to detect attacks
on the IoT networks only partially cover the large set of protocols and communication
technologies that are typically used in smarthomes. These solutions also require
the development of specific probes for each relevant protocol, or necessitate a
considerable reconfiguration effort in case of wireless network evolution, e.g. change of the
WiFi frequency \cite{Cordeiro_802.11ad_2010}.

In this paper, we propose RadIoT, a novel solution that is aimed 
at filling this gap, and that is designed to complement existing solutions targeting traditional protocols such as WiFi. The proposed approach is protocol agnostic and can cover all the communication protocols available in the
target IoT network. It consists in monitoring and profiling the radio activities 
at the physical layer, especially by observing the power and time of 
transmission. Moreover, this solution requires only a light learning phase, without impacting
the objects. 

\section{Threat model}
\label{chap:threat}

In a smarthome context, the attacker can proceed remotely (e.g. without physically interacting
with the devices) either i) from the
Internet or outside the smart place, or ii) within the smart place. Its objectives may be to:
\begin{itemize}
    \item modify the devices to prepare a home intrusion
            \cite{Ho_SmartLocksLessons_2016};
    \item collect confidential information within the devices
            \cite{Newlin_MouseJackKeySnier_2016};
    \item control devices in order to bounce elsewhere
            \cite{Ronen_IoTGoesNuclear_2018};
    \item control devices to affect their functionality
            \cite{ronen_extended_2016}.
\end{itemize}

To compromise the targeted devices, the attacker can either
use his own devices or use an already compromised
one.
In both situations, the attack will result in malicious interactions with the devices of the smarthome, that may occur concurrently with the legitimate interactions.

In our study, we focus on the monitoring of the radio-activities generated by such interactions, considering both the time and the location of the activities. 
A malicious behavior corresponds to the detection of a radio activity corresponding to one of the following situations:
\begin{enumerate}
    \item from a location that is never used by users;
    \item using a wireless technology never used by users;
    \item during an unusual period of the day;
    \item following an unusual scenario (e.g., switch on lights without disabling the alarm first);
    \item with unusual intensity.
\end{enumerate}

According to Section \ref{chap:related}, most of existing solutions focus on the monitoring of the communications through a central gateway in the smarthome, and investigate in particular the traffic at the 
network layer. Hence, they are likely to be inefficient to detect the aforementioned malicious
behaviors. 
In the next section, we present our proposed approach that provides a complementary protection mechanism and a better coverage of potential malicious activities targeting smarthomes.
\section{Approach overview}
\label{chap:overview}

To provide protection against the threat scenarios presented in Section \ref{chap:threat}, a solution could be to develop a specific probe for each communication protocol used in the smarthome, and to monitor the corresponding network traffic. However, this solution would be not practically feasible as it requires the availability of the specification of all the protocols used in the smarthome, which is not always possible and it would be costly. Also, it would be not suitable for dynamic environments characterized by the evolution of the wireless IoT technologies used, e.g. when new devices are introduced. 

Our solution named RadIoT monitors the radio activities on frequencies commonly used by wireless communications. It considers only the received physical signal intensity, without relying on the specification of the protocols. Also it does not require any modification of the existing devices.

Two main phases can be distinguished. The first one deals with the creation of a reference model that represents the radio-activities generated by the normal use of the connected devices in the absence of malicious activities. The second phase corresponds to the detection of \textit{illegitimate behaviors} or
\textit{misbehaviors}. It consists in 1) the real-time capture of radio activities in the smarthome,
2) the comparison of the captured data with the reference model as explained in the section \ref{subsec:detectres}, and 3) the generation of alerts whenever a deviation from the reference model is detected.

\begin{figure}[ht]
    \centering
    \includegraphics[width=0.75\columnwidth]{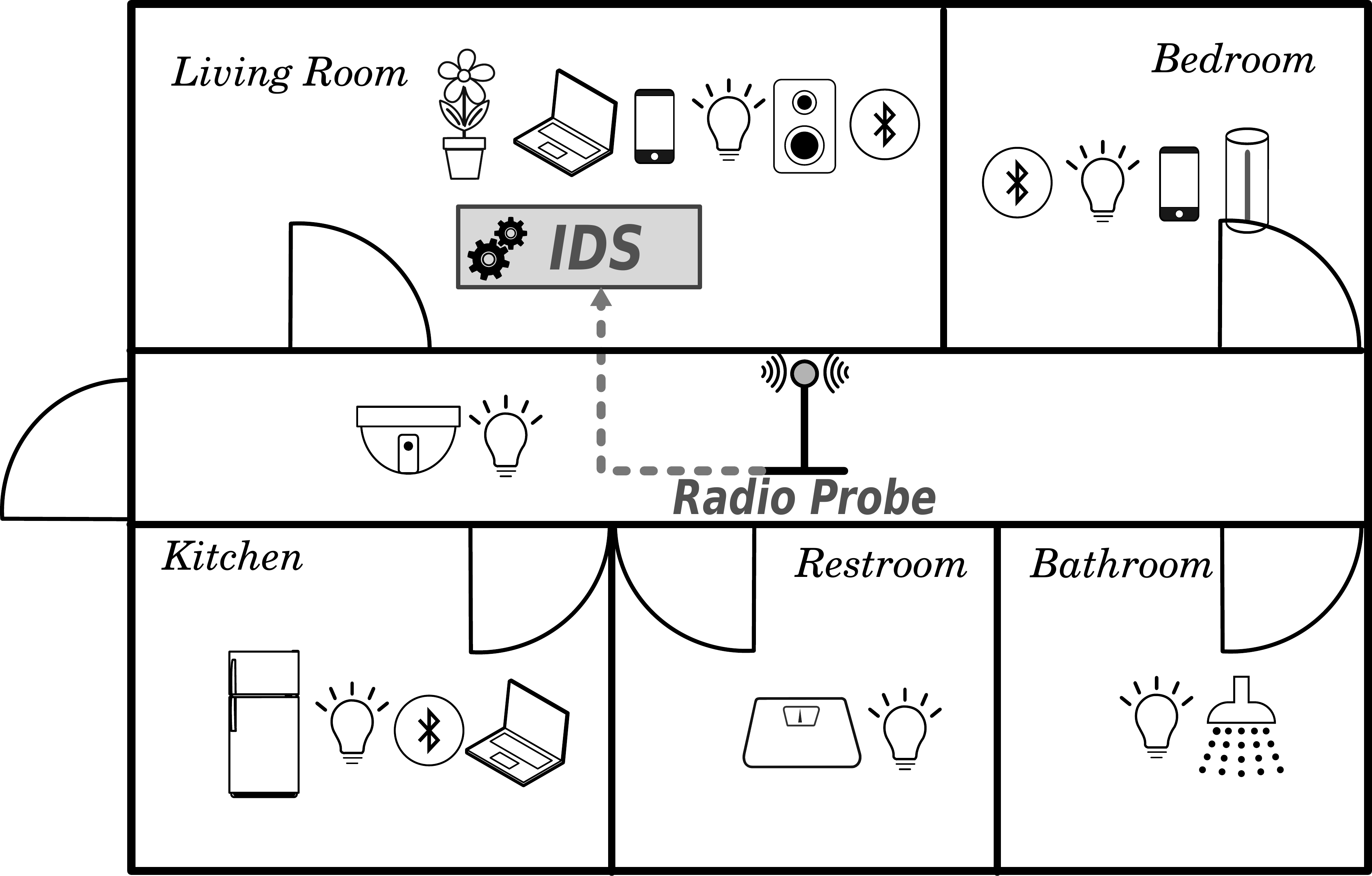}
    \caption{A smarthome example integrating the IDS and one radio probe}
    \label{fig:archi}
\end{figure}

As illustrated in Figure \ref{fig:archi}, RadIoT relies on two main
components: 1) one or several \textit{Radio Probes} deployed in strategic locations of the home, and 2) the \textit{IDS} that processes the data received from the probes and implements the detection models.

The probes are based on \textit{Software-Defined Radio (SDR)} technology and capture the radio activities on specific frequencies. These activities are characterized by the signal power, the reception time and the frequency used. The frequencies are specified taking into account the deployed devices, and can easily be adapted to take into account new protocols.
\section{Radio Probe}
\label{chap:design}

The \textit{Radio Probe} must be easily deployed in a smarthome. In addition, it must fulfill the following requirements:
\begin{itemize}
    \item Capability to monitor activities on different bandwidths;
    \item Low cost
    \item Easy configuration and installation with the help of a security expert;
    \item Portability;
    \item Secure connection to the IDS;
    \item Non-invasiveness, i.e. no modification of IoT devices.
\end{itemize}

These requirements can be fulfilled with a \textit{Software-Defined-Radio (SDR)} peripheral monitoring radio signals on a large frequency bandwidth. Accordingly, the probe can easily observe communications on traditional protocols (Zigbee, Bluetooth, 868MHz, etc.) with no reconfiguration.

\begin{figure}[ht]
   \centering
    \includegraphics[width=0.92\columnwidth]{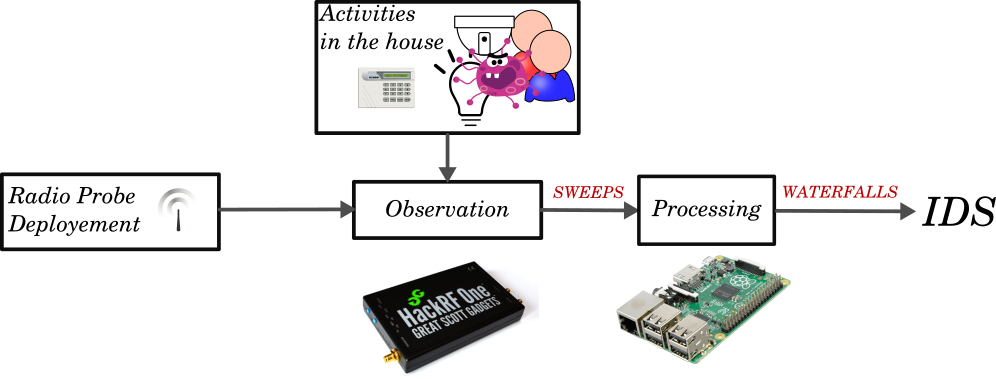}
    \caption{Radio activities monitoring}
    \label{fig:monitor}
\end{figure}

As illustrated in Figure \ref{fig:monitor}, two main tasks can be distinguished: the \textit{observation} task consists in capturing the radio activities on relevant frequencies, and the \textit{processing} task consists in formatting the raw data and extracting information to be used by the IDS to elaborate the detection models. 

\begin{figure}
    \centering
    \includegraphics[width=0.88\columnwidth]{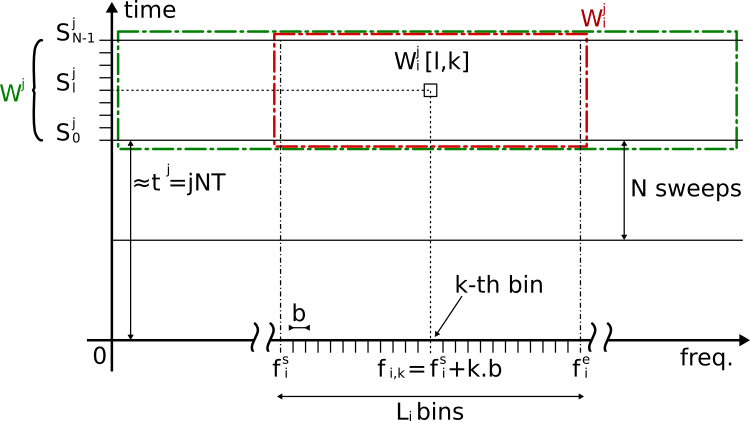}
    \caption{Waterfall notation}
    \label{fig:notation}
\end{figure}

We measure the radio-communication power (in dBm) associated to the monitored frequencies. Figure \ref{fig:notation} summarizes the main concepts and notations illustrating the power measurement process and the corresponding outputs.

Relevant frequencies to be monitored by the probe correspond to discontinuous ranges (e.g. [800-900MHz] and [400-500MHz] generally used for home-automation and [2.4-2.5GHz] used for WiFi, Zigbee etc.) \cite{al-fuqaha_internet_2015}. In the following, these
frequencies will be represented as a set $F$ of $M$ frequency ranges
$F_i=(f^s_i,f^e_i)$, between $f^s_i$ and
$f^e_i$~KHz, ($M=|F|$). The measurement of the power associated to these frequencies implies the use
of Fast Fourier Transforms (FFT). Let us denote $b$, the bin width, in KHz, which corresponds to
the frequency sampling resolution of the FFT. 

Depending on the SDR technology, several FFTs may be needed to cover the
relevant frequency ranges (e.g., HackRF is able to compute
the FFT on a 20MHz bandwidth). The measurement of the power associated to $F_i$ at a given time stamp is called a
\textit{sweep} and the resulting vector is noted $S_i$. This vector contains
$L_i=\frac{f^e_i - f^s_i}{b}$ values, corresponding to the powers measured for each bin of the FFT. The rate at which this sweep
is achieved depends on the SDR used. Let us denote by $T$ the time in seconds between two consecutive sweeps. 

\begin{figure}[ht]
    \centering
    \includegraphics[width=0.9\columnwidth, height=0.6\linewidth]{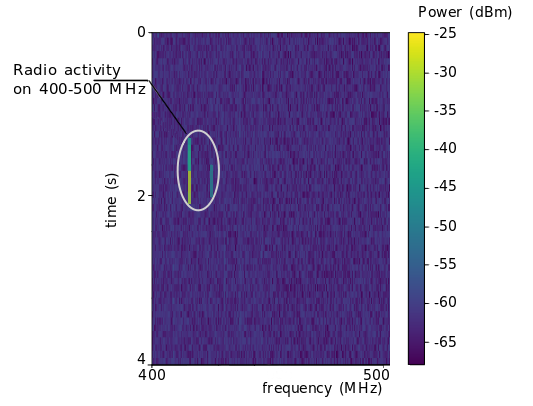}
    \caption{Waterfall example}
    \label{fig:waterfall}
\end{figure}

The SDR is configured to calculate the sweeps for all the $M$ considered frequency ranges. This process is
repeated continuously. The results of $N$ consecutive sweeps are aggregated together to form a \textit{waterfall} noted $W$.
The j-th waterfall obtained with the SDR is noted $W^j$. In the following, the $j$ superscript will be omitted.
Let us d   enote by $W_i$ the subset of the waterfall associated to the frequency range $(f^s_i,f^e_i)$.
Each waterfall is time-stamped with the date $t$ of its first sweep: $(t, W)$. Hence, a waterfall
is a matrix with a size of $\sum_i{L_i}\times N$. It is usually displayed as illustrated in figure
\ref{fig:waterfall}. This figure represents a waterfall composed of
$100$ sweeps (with $T=$0.0375~seconds) that was measured by a HackRF peripheral configured for a frequency range 400~MHz to 500~MHz. The horizontal index $k$ corresponds to the k-th bin and the vertical index
$l$ to the l-th sweep. Thus, the pixel at position $(l, k)$ corresponds to the power
$W_i[l, k]$, observed at time $t+l*T$, associated to the frequency $f^s_i+k*b$~KHz. The scale on the right of
the figure maps the pixel color to the measured power: yellow
corresponds to a high power signal while dark-blue corresponds to a low power signal.

The power measurements are stored using the type \textbf{float-64}, then the size of each waterfall is $8*N*\sum_i{L_i}$ and they are generated at speed of $8*\frac{\sum_i{L_i}}{T}$ bytes per seconds. It is noteworthy that the IDS can be connected to several probes, hence the outputs are tagged with the probe id.
\section{Model Generation}
\label{chap:model}

The model implemented in the \textit{IDS} to detect attacks is based on machine learning techniques widely used to implement anomaly-based IDS. During the learning phase, the waterfalls collected through the radio probes reflect legitimate activities in the smarthome and are used to calibrate the model. This section details the model generation as well as the learning algorithm chosen, 
which is the autoencoder. More information is provided about the features used as inputs of the autoencoder and how these
are extracted from the waterfalls.

\subsection{Model selection}

Two main characteristics of the collected data are relevant to guide the IDS model design decisions. First, the probes continuously collect during the learning phase the radio-activities at a fine granularity, generating a large amount of waterfalls that need to be pre-processed to reduce the data size. 
Second, all waterfalls are associated to only one class of behavior: the legitimate one.
The problem addressed in our context is to detect an anomaly in the current observation in comparison with a reference model. In
machine learning, this problem is defined as a classification problem. Several classification algorithms are available such as a Neural Networks, K-NN, Support Vector Machines (SVM), etc. The selection of the most suitable algorithm depends on the data collected. For example, Y. Sung \cite{sung_intelligent_2016} investigates the use of Bayesian networks to detect intruders within a smart environment, through the use of RSSI in Bluetooth beacon. E. Hodo and al. \cite{hodo2016} presents an IDS based on artificial neural networks (ANN), able to detect DDoS/DoS attacks on IoT networks, with 99.4\% accuracy against simulated attacks.

Our problem is related to a one-class machine learning classifier that uses only positive examples. 
In addition, due to the large size of the collected data (see section \ref{chap:design}) and in order to achieve reasonably low processing times, data dimension reduction is required to reduce the number of random variables under consideration (i.e. features). For the results presented in the following, we use an autoencoder neural network that provided satisfactory results in similar situations \cite{jsan6040032}. The comparative analysis of different machine learning algorithms is planned for future work.

\subsection{Autoencoder neural network}

An autoencoder is a deep neural network \cite{ng:autoencoder} that is trained to produce outputs that are identical to its inputs.
It is structured as a non-recurrent feedforward neural network with the same number
of inputs and outputs, and with an odd number of hidden layers including a bottleneck. The bottleneck
allows to learn a compressed representation of the
inputs that enables the reconstruction of the outputs from these inputs. 
An autoencoder is composed of two parts.
The first part encodes the inputs $x$ onto the first layers until the bottleneck, as a function $f(x)$.
The second part is composed of the other layers that represent a function $g(y)$ aimed at expanding the dimension of $y$. The overall goal is to learn these
two functions so that $g(f(i)) \approx i$ where $i$ corresponds to the inputs.
The learning process of these two functions relies on the minimization of the loss function $L(x,g(f(x)))$
where $L$ can correspond for instance to the mean squared error.

\subsection{Features extraction}

To generate a representative model of the legitimate radio communications, the autoencoder algorithm requires the definition of relevant \textit{features} used as inputs. These \textit{features} are extracted
from the waterfalls generated from the data collected by the probes. The entire waterfall $W$ or any $W_i$ or any sub-bandwidth of these $W_i$ can be used.
More precisely, each waterfall is split according to the bandwidth or to the relevant frequencies considered. If necessary, a given bandwidth can be split into overlapped slices. So, a waterfall can be split into ZigBee channels and into WiFi channels.
The features can be based on the main metrics that characterize the signal
power on the selected communication bandwidths: maximum, minimum, mean, median, standard deviation and sum.
Hence, considering a range of frequencies $k_s..k_e$ of the waterfall $W$, these metrics are formally defined as follows:

$\Omega(W,k_s,k_e)=\{~W[k,l] ~|~ k\in [k_s..k_e], ~l\in[0..N[~\}$

$\mathrm{Max}(W,k_s,k_e)=\max~\Omega(W,k_s,k_e)$

$\mathrm{Min}(W,k_s,k_e)=\min~\Omega(W,k_s,k_e)$

$\mathrm{Mean}(W,k_s,k_e)=\mean~\Omega(W,k_s,k_e)$

$\mathrm{Median}(W,k_s,k_e)=\median~\Omega(W,k_s,k_e)$

$\mathrm{Std}(W,k_s,k_e)=\std~\Omega(W,k_s,k_e)$

$\mathrm{Sum}(W,k_s,k_e)=\sum~\Omega(W,k_s,k_e)$
    
Additional features are included to take into account the time stamp of the waterfall as explained in \cite{July31_Encodingcyclicalcontinuous_2018}.
Then, in order to learn the radio-activities patterns evolution over time, the features extracted from the ten last waterfalls also used as inputs in the autoencoder, considering a sliding window. In total, 80 features are used for the training of the autoencoder.

The autoencoder takes these features as inputs. By using backpropagation algorithm, it is fitted to match the legitimate behaviors. Then, during the monitoring and detection phase, this model is used as the reference model. 
During operation, the IDS extracts the features for each waterfall and sends
the corresponding vector to the autoencoder. If the outputs show significant deviations from the input
vector, i.e. more than a fixed threshold $S$, an anomaly is reported to the users. The threshold $S$ is calculated empirically through an
experimentation phase presented in the next section.
\section{Experimentation}
\label{chap:expe}
This section describes some preliminary experiments we have carried out in order to assess our intrusion detection 
approach in a inhabited smarthome, purposely equipped with various
connected objects. This environment is illustrated in
figure \ref{fig:archi}. Various connected devices are deployed in the smarthome using different heterogeneous protocols (Bluetooth, ZigBee, WiFi and 433MHz) listed in Table \ref{table:obj}.
\begin{table}[ht]
\caption{Deployed IoT devices and their characteristics}
\centering
\begin{tabular}{p{0.1cm}p{2cm}p{3cm}p{1cm}p{0.8cm}}
ID & Name & Aim & Protocol & Number \\ \hline
\hline 
1 & Netatmo Sensors & Air Ambiant Sensor & WiFi & 1 \\
2 & D-Link Camera & Live Streaming Video & WiFi & 1 \\
3 & Phones, Laptops & General Internet Services & WiFi & 4 \\
4 & Parrot Flowerpot & Flowerpot & Bluetooth & 1\\
5 & Terraillon scale & Scale & BLE & 1\\
6 & Philips Hue & Lightbulb & ZigBee & 2\\
7 & Hoover Neato & Hoover & WiFi & 1\\
8 & Domotic sensor & Simulated & 433MHz & 1
\end{tabular}
\label{table:obj}
\end{table}

For this preliminary study aimed at exploring the feasibility of the proposed approach, a single probe was installed in the smarthome to capture the radio activities generated by the IoT devices. The first experiment consisted in capturing the radio activities generated by the devices without deliberately injecting attacks. These data were used as training set to validate a first reference model. In a second phase, we set another experiment with a new reference model established the same way. Then some attacks were carried out simultaneously with legitimate behavior in an attack dataset. This data were used to assess the detection performance of the IDS. 
The following subsections provide some details about the elaboration of the model implemented by the IDS and the datasets used for its training and testing. Information is also provided about the generation of the attack data as well as the modus operandi of injection. Finally, the results are presented and discussed.

\subsection{Data collection and RadIoT model training}

Our probe is a HackRF One device, connected to a Raspberry Pi 3 equipped with a hard drive to store the captured data\footnote{https://github.com/mossmann/hackrf/wiki}.
For this experiment, we focused only on three bandwidths, 400-500 MHz, 800-900 MHz and 2.4-2.5 GHz,
which cover most of the communications usually used in IoT, using a low cost antenna. Moreover, the two hackrf\_sweep parameters "number of samples" and "bin width", that control the FFT resolution, are set to 8192 and 200 KHz respectively.

Two different datasets were collected. 
The first experiment consisted of 300GB of radio activities recorded continuously during 2 consecutive weeks, without any attack activity, to validate and calibrate the autoencoder. The second experiment consisted in collecting a training/testing dataset of 7 days with an attack activity of 2 days with different attacks covering the three selected frequency bandwidths. 

As regards the training and testing of a reference model, 70\% of the dataset is used to set a reference model and the remaining 30\% for testing and validating this model. 
For both experiments, the datasets are processed to extract the features as described in Section \ref{chap:model}. More precisely, each waterfall is split into 400-500 MHz, 800-900 MHz and 2.4-2.5 GHz slices and one features vector is generated for each slice.

The autoencoder models are generated with TensorFlow library\footnote{https://www.tensorflow.org/}. For both experiments, the  autoencoder is composed of 3 hidden layers respectively with 87.5\%, 75\% and 87.5\% neurons on each. Two activation functions have been tested: {\em softplus} and {\em sigmoid}. {\em softplus} provides better results.
A reference model is then defined for each bandwidth considered.  

\subsection{RadIoT attacks campaign}
\label{sub:attack}

The attack injection campaign is composed of two steps: 
1) attack definition and simulation; 2) attack injection.

\subsubsection{Attack definition and simulation}
\begin{figure}[t]
    \centering
    \includegraphics[width=0.70\columnwidth]{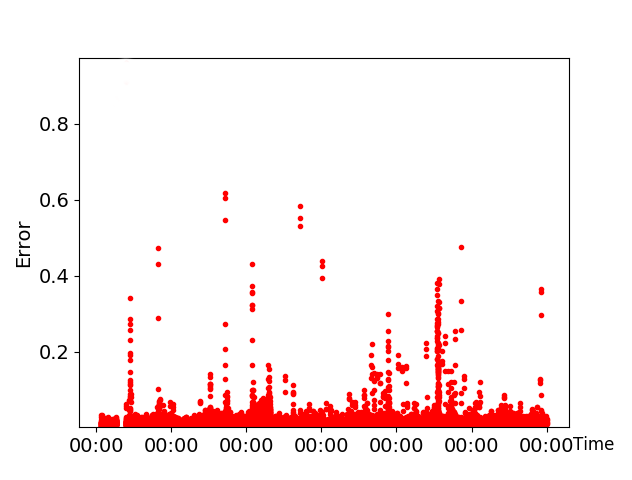}
    \caption{Test error - dataset 2 at 800-900MHz}
    \label{fig:resultstest}
\end{figure}
\begin{figure}[t]
    \centering
    \includegraphics[width=0.70\columnwidth]{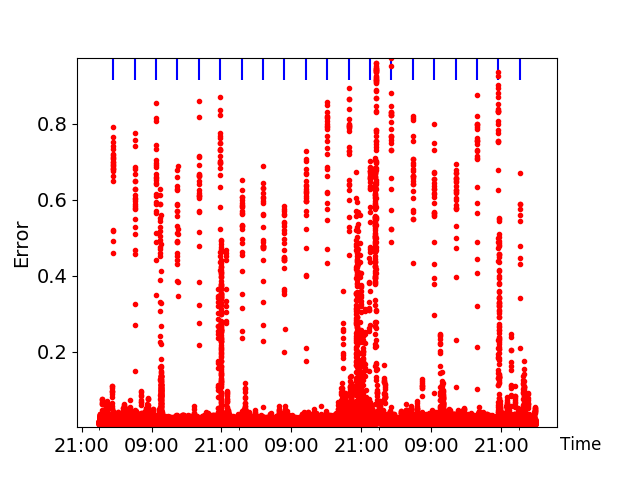}
    \caption{Attack error - dataset 2 at 800-900MHz}
    \label{fig:resultsattack}
\end{figure}

As our detection approach only relies on the analysis of radio activities, it is not necessary to perform real attacks, including their exact and complete payloads, but it is sufficient to simulate these attacks by generating radio activity samples emulating the corresponding behaviour. An attack is defined by: 1) the intensity of the signal power for the selected frequencies considering three levels (High, Normal or Low) and 2) its duration (Long, or short). An additional parameter is also defined to specify the time between consecutive attacks. The definition of the considered attacks was based on the knowledge collected from published IoT attacks.

We selected a set of eight attacks described in the table \ref{table:attacks}. Some of them are performed with a generic signal transmitter such as the HackRF device (such as 868 MHz attack), and some are performed with dedicated dongles (such 
as Wifi Dos attack or Rogue Access Point (AP) Attack \cite{Waliullah_WirelessLANSecurity_2014}). 
For each attack, Table \ref{table:attacks} provides the information about the intensity and the duration of the corresponding signal. Let us note that DOS attacks (1 and 8) are simulated by generating a high power message followed by a simulated shutdown of the legitimate AP. 
This set is then injected in the smarthome experimental environment during several days, to represent the different situations of malicious behavior described in Section \ref{chap:threat}. The parameters used to calibrate the intensity of the power signal and the attack duration for the different attack categories are defined empirically.
\begin{table}[!ht]
\caption{Set of injected attacks}
\centering
\begin{tabular}{p{0.07cm}p{0.7cm}p{2.1cm}p{0.9cm}p{1.6cm}p{1.3cm}}
ID & Protocol & Type & Intensity & Duration & Frequency \\ \hline
\hline
1 & WiFi & DoS & High & Long (20min) & 2430 \\
2 & WiFi & Deauthentification & Normal & Short (1min) & 2437 \\
3 & WiFi & Rogue AP & Normal & Long (4min) & 2412 \\
4 & BLE & Man in the Middle & Normal & Long (4min) & 2400-2500 \\
5 & Zigbee & Fake association & Normal & Short (1min) & 2470 \\
6 & Zigbee & Fake data send & Normal & Long (4min) & 2470 \\
7 & 868MHz & Simulated & High & Short (1min) & 868 \\
8 & 433MHz & DoS & High & Long (10min) & 433
\end{tabular}
\label{table:attacks}
\end{table}

\subsubsection{Attack injection}

The attacks are injected with an attack generator tool that was purposely 
developed for the experiments (due to space limitation, the description of the tool is out of the scope of
the paper).
It provides an easy-to-use interface to configure and to perform a particular set of attacks. It records the time, called $tatt_{send}$, of the attack generation.
20 attack campaigns were carried out. Each campaign is
3 hours and 40 minutes long, with 1 hour separating two consecutive campaigns. Each campaign includes 6 attacks (attack 2 to 7), the duration of each attack depends on the type of attack (see Table \ref{table:attacks}), with 20mn separating two consecutive attacks. The DOS attacks (attack 1 and 8) are performed only once at a different period from the others.

\subsection{Detection results}
\label{subsec:detectres}

To evaluate the detection performance of the IDS, traditional metrics are computed: False Positives ($FP$), True Positives ($TP$), False Negatives ($FN$), True Negatives ($TN$). These metrics are used to calculate $Precision$, $Recall$ and $True~Negative~Rate~(TNR)$ defined as:
\begin{gather*}
Precision=\frac{TP}{TP + FP} \hspace{1cm}
Recall=\frac{TP}{TP+FN}\\
TNR=\frac{TN}{TN+FP} 
\end{gather*}

These metrics are evaluated on the attack dataset collected as explained in subsection \ref{sub:attack}. To compute them, it is necessary to measure the error between the inputs and the outputs of the autoencoder, thanks to the following equation:
\begin{equation*}
Error(i)=output_i - input_i, i \in\{1,80\}
\end{equation*}

Then, we implemented the detection method based on the assessment of the likelihood of the observed error vector including all the considered features. The likelihood is computed based on the concepts described in \cite{luvison2009} where the error probability for each feature is approximated by a Gaussian distribution.
Then, an anomaly is notified when the probability associated to the error vector exceeds a given threshold $t$. The threshold is estimated empirically based on the analysis of the $TNR$ measured on the testing dataset. 

In the first experiment, the first training dataset was used to evaluate whether a model can correctly fit two weeks of communications on three bandwidths. When the error rate is sufficiently low, a.k.a. 0.1\% of the maximum observed error, it is considered that the model correctly fits the data. 
Next, the corresponding testing dataset was used to determine how to calculate the threshold to detect an anomaly. We calculated the likelihood of the observed errors and evaluated the $TNR$ for different threshold values until there is only few or no $FP$. For instance, for the 800-900MHz bandwidth, a threshold of 0.6 is perfect to have no $FP$ on the testing data.
The second experiment was used to determine the detection performance of the IDS on the attack dataset. The threshold was set based on the corresponding testing data and the metrics were evaluated with that threshold. Table \ref{tab:results} describes the results for each bandwidth.

\begin{table}[b]
    \centering
    \caption{Results for each bandwidth}
    \begin{tabular}{|p{1.3cm}p{1.1cm}|p{1.1cm}|p{1cm}|p{1cm}p{1cm}|}
         & & & Testing & \multicolumn{2}{c}{Attack} \vline\\
        Bandwidth (MHz) & ID Attack & Threshold & TNR & Precision & Recall \\ \hline  \hline
        400-500 & 8 & 0.3 & 99.40\% & 99.83\% & 100.00\% \\
        800-900 & 7 & 0.6 & 99.60\% & 82.76\% & 92.31\% \\
        860-870 & 7 & 0.2 & 99.80\% & 100.00\% & 96.15\% \\ 
        2400-2500 & 1,2,3,4,5,6 & 0.3 & 98.61\% & 83.33\% & 1.15\% \\
        2410-2420 & 3 & 0.3 & 99.00\% & 98.99\% & 79.89\% \\
        2400-2410 & 4 & 0.3 & 97.01\% & 98.55\% & 57.46\% \\
        2465-2475 & 5,6 & 0.4 & 98.01\% & 10.00\% & 1.67\% \\
        2430-2440 & 1,2 & 0.6 & 99.60\% & 100.00\% & 4.00\% \\
        2420-2430 & 1 & 0.4 & 99.60\% & 100.00\% & 100.00\% \\
    \end{tabular}
    \label{tab:results}
\end{table}

Figures \ref{fig:resultstest} and \ref{fig:resultsattack} present the error plot on the 800-900 MHz bandwidth for the testing and the attack data respectively. The blue marks on figure \ref{fig:resultsattack} identify the starting dates of the attacks. For this example, the error patterns computed on the attack data show significant deviations compared to the test dataset.
In presence of attacks, as shown in figure \ref{fig:resultsattack}, the error is far superior than the highest error computed from the testing dataset (depicted in figure \ref{fig:resultstest}).

Considering Table \ref{tab:results} results, it is shown that for the 800-900 MHz and 400-500 MHz bandwidths, almost all attacks are detected and only a few false alarms are raised. Moreover, as 800-900MHz bandwidth is shared by many protocols, better detection performance are obtained when the analysis is done at a finer granularity, e.g., by focusing only on the bandwidth used by the attack (e.g., 860-870MHz for the 7\textsuperscript{th} attack). The results improve significantly as the IDS does not raise any False Positive and only one attack is not detected.

Not surprisingly, the results for the 2400-2500MHz bandwidth are mixed and show a higher variability. As this bandwidth is used by several protocols (similarly to 800-900MHz), we created several models at a finer granularity, for each 10MHz (e.g. 2400-2410, 2405-2415, etc.). The results improve significantly: all the Rogue AP and DoS attacks are detected, as well as more than half of the BLE attacks (57.46\%). However, ZigBee attacks and the WiFi deauthentification attack are not well detected. We believe that this is partly due to our hardware (antenna, HackRF) which is not optimized for this bandwidth. Moreover, the 2\textsuperscript{nd} attack is short and uses the same channel as the legitimate AP, therefore, probably these attacks are too short to be easily detected. A longer learning phase should improve the results and we plan to validate this assumption in future work.
An interesting behaviour was observed about the IDS detection performance when Denial-of-Service attacks (i.e., 1 and 8) were performed. The attack signal itself is not well detected, especially on the 2400-2500 bandwidth. However, the effects of these attacks are clearly recognizable, as the communications suddenly stop and some devices start to send many messages during a short period of time, and some beacons are no more received. Moreover, when the DoS stops, the AP restarts and potentially changes its communication channel. As a consequence, the radio-communication activities observed after the DoS attack exhibit different patterns in particular on the bandwidth associated to the new channel (in our experiment 2400-2420 bandwidth), compared to the learned reference model. In our experiment, we consider that this effect corresponds to an attack, as this behavior follows an illegitimate action.
\section{Limitations \& Future Work}
\label{chap:limits}

The first results show that RadIoT is able to detect most of the attacks that we have carried out within a smarthome, even during high periods of activity by the users. However, a few attacks were not detected. In this work, the experimentation was carried out with only one probe. However, it is very difficult for a single probe to catch all the traffic and detect an attack performed from outside or behind a load-bearing wall. We believe that the performance of our IDS will be significantly improved by deploying more probes and correlating the activities between them. Experiments are planned in the future to validate this assumption. For instance, to detect more precisely an illegitimate source location of a signal, the difference between the power received by different probes can be learned. Therefore, the correlation between activities monitored on different probes should provide better discrimination between legitimate behaviors and attacks. Another possible area of improvement would to use a more sensitive and precise antenna than the one used in our experiments to better detect attacks with a low power intensity. Moreover, an analysis has to be performed to correctly consider all the potential physical phenomena that can impact our IDS, such as the change of the communication channel example.

Our model shows interesting results but some sensitivity studies still have to be done. More precisely, more work on the definition and experimentation of the activation functions, the autoencoder architecture and features selections have to be considered and may
improve the detection performance of the solution. 
It could also be interesting to build an adaptive model that can learn continuously and adapt itself to the changes that occur in a smarthome, for example a new device or a big change in the users behaviour. Moreover, at this stage we did not consider the difference between the communications behaviors during the week and the week end, this should improve significantly the learning of the legitimate behavior.
\section{Conclusion}
\label{chap:concl}

This paper presents a new intrusion detection solution called RadIoT to protect connected areas like smart homes and smart factories. It is based on the monitoring and the anomaly detection of radio activity through neural networks. The first experiment has been carried out in a smarthome equipped with various connected objets. RadIoT shows relevant results (88.46\% Precision and Recall) for bandwidths that are not usually covered by traditional IDS even though they are frequently used in smarthome environment, e.g. 868MHz and 433MHz. Moreover, it is also able to detect some attacks on IoT protocols (Rogue AP, BLE attacks) with only a low configuration and a low-cost prototype, and with only one antenna for all protocols. As future work, we plan to continue the experimentation and use multiple probes, and also validate our solution in other smart environments (such as a smartfactory) which can be more or less constrained compared to smarthome. We also plan to release the data and share it with the scientific community to test other machine learning techniques or algorithms.

\bibliography{IEEEabrv,Biblio/biblio3}
\bibliographystyle{IEEEtran}

\end{document}